\documentclass[sigconf]{acmart}
\usepackage{mathtools}
\usepackage{multicol, multirow}
\usepackage{booktabs}
\usepackage{amsmath}
\usepackage{amsfonts}
\DeclareMathOperator*{\argmax}{arg\,max}
\DeclareMathOperator*{\argmin}{arg\,min}
\usepackage{caption}
\usepackage{bbding}
\usepackage{enumitem}
\usepackage{subfloat}
\usepackage{graphicx}
\usepackage{xcolor}
\usepackage{subfig}
\usepackage[ruled,linesnumbered]{algorithm2e}
\usepackage{threeparttable}
\usepackage{bm}
\copyrightyear{2024}
\acmYear{2024}
\setcopyright{acmlicensed}\acmConference[CIKM '24]{Proceedings of the 33rd ACM International Conference on Information and Knowledge Management}{October 21--25, 2024}{Boise, ID, USA}
\acmBooktitle{Proceedings of the 33rd ACM International Conference on Information and Knowledge Management (CIKM '24), October 21--25, 2024, Boise, ID, USA}
\acmDOI{10.1145/3627673.3679712}
\acmISBN{979-8-4007-0436-9/24/10}

\begin{document}

\title{OptDist: Learning Optimal Distribution for Customer Lifetime Value Prediction}
\author{Yunpeng Weng}
\authornotemark[1]
\email{edwinweng@tencent.com}
\orcid{0000-0001-7593-2169}
\affiliation{%
  \institution{FiT, Tencent}
  \city{Shenzhen}
  \state{Guangdong}
  \country{China}
  \postcode{518057}
}

\author{Xing Tang}
\authornote{Contributed equally}
\email{xing.tang@hotmail.com}
\orcid{0000-0003-4360-0754}
\affiliation{%
  \institution{FiT, Tencent}
  \city{Shenzhen}
  \state{Guangdong}
  \country{China}
  \postcode{518057}
}

\author{Zhenhao Xu}
\email{zenhaoxu@tencent.com}
\orcid{0009-0002-2348-2148}
\affiliation{%
  \institution{FiT, Tencent}
  \city{Shenzhen}
  \state{Guangdong}
  \country{China}
}

\author{Fuyuan Lyu}
\authornote{This work was done when working at FiT, Tencent.}
\email{fuyuan.lyu@mail.mcgill.ca}
\orcid{0000-0001-9345-1828}
\affiliation{%
  \institution{McGill University \& MILA}
  \city{Montreal}
  \country{Canada}
}

\author{Dugang Liu}
\authornotemark[3]
\email{dugang.ldg@gmail.com}
\orcid{0000-0003-3612-709X}
\affiliation{%
  \institution{Guangdong Laboratory of Artificial Intelligence and Digital Economy (SZ)}
  \city{Shenzhen}
  \state{Guangdong}
  \country{China}
  \postcode{518057}
}

\author{Zexu Sun}
\authornotemark[2]
\email{sunzexu21@ruc.edu.cn}
\orcid{0000-0002-6727-6242}
\affiliation{%
  \institution{Renmin University of China	}
  \city{Beijing}
  \country{China}
}

\author{Xiuqiang He}
\authornote{Corresponding authors}
\email{xiuqianghe@tencent.com}
\orcid{0000-0002-4115-8205}
\affiliation{%
  \institution{FiT, Tencent}
  \city{Shenzhen}
  \state{Guangdong}
  \country{China}
  \postcode{518057}
}

\renewcommand{\shortauthors}{Yunpeng Weng et al.}
\begin{abstract}
Customer Lifetime Value (CLTV) prediction is a critical task in business applications, such as customer relationship management (CRM), online marketing, etc. Accurately predicting CLTV is challenging in real-world business scenarios, as the distribution of CLTV is complex and mutable. Firstly, there is a large number of users without any consumption consisting of a long-tailed part that is too complex to fit. 
Secondly, the small set of high-value users spent orders of magnitude more than a typical user leading to a wide range of the CLTV distribution which is hard to capture in a single distribution. 
%Secondly, the small set of high-value users spend orders of magnitude more than a regular user. This leads to a wide range of the CLTV distribution which is hard to capture in a single distribution.
Existing approaches for CLTV estimation either assume a prior probability distribution and fit a single group of distribution-related parameters for all samples, or directly learn from the posterior distribution with manually predefined buckets in a heuristic manner. 
However, all these methods fail to handle complex and mutable distributions. 
In this paper, we propose a novel optimal distribution selection model (\textbf{OptDist}) for CLTV prediction, which utilizes an adaptive optimal sub-distribution selection mechanism to improve the accuracy of complex distribution modeling. Specifically, OptDist trains several candidate sub-distribution networks in the distribution learning module (DLM) for modeling the probability distribution of CLTV. Then, a distribution selection module (DSM) is proposed to select the sub-distribution for each sample, thus making the selection automatically and adaptively. Besides, we design an alignment mechanism that connects both modules, which effectively guides the optimization. We conduct extensive experiments on both two public and one private dataset to verify that OptDist outperforms state-of-the-art baselines. Furthermore, OptDist has been deployed on a large-scale financial platform for customer acquisition marketing campaigns and the online experiments also demonstrate the effectiveness of OptDist. 
\end{abstract}

%The source code of the model implementation is available\footnote{https://anonymous.4open.science/r/CLTV-8F8F/}.

\maketitle
\section{Introduction}

Predicting customer lifetime value (CLTV) is a task that refers to the estimation of potential revenue a user may bring to a platform or company~\cite{wang2019deep,li2022billion}. The accurate prediction of CLTV holds significant importance in various commercial settings, such as online advertising, marketing campaigns, and customer retention strategies~\cite{vanderveld2016engagement,esdpm,rerum}. For example, CLTV is helpful for further decision-making in customer acquisition marketing campaigns with resource constraints. Specifically, we can predict CLTV for users in a specified duration on the commercial platform and put more resources into attracting high-value customers, leading to efficient and effective utilization of budgets and higher Return on Investment (ROI).

Some conventional methods can already be directly used for CLTV modeling. One is the \textit{statistic-based} approach~\cite{fader2009probability,fader2005rfm,fader2005counting,colombo1999stochastic}, which assumes a probability distribution for CLTV and obtains the parameters of the statistical distribution based on historical statistical data, such as each customer's consumption frequency and so on. 
With the advance of deep learning, some \textit{value-based} approaches utilize a neural network to predict the exact value of CLTV~\cite{xing2021learning,wang2019deep,chen2018customer,zhao2023percltv}.
However, \textit{statistic-based} approaches only rely on users' history statistics without consideration of personalization. For example, historical statistics such as frequency and recency may be the same for some users, so these methods can only roughly predict the same CLTV for them. 
As to \textit{value-based}  approaches, most of them adopt the Mean Squared Error (MSE) as the loss function to train a regression model, which is very sensitive to the outliers in CLTV. This leads to instability in the training and degradation of the prediction performance. Therefore, these approaches cannot predict CLTV well due to the complex and mutable distribution.

\begin{figure}[!t]\centering    
	\subfloat[Cluster \#1] {
		\label{fig:a}
		\includegraphics[width=0.48\columnwidth]{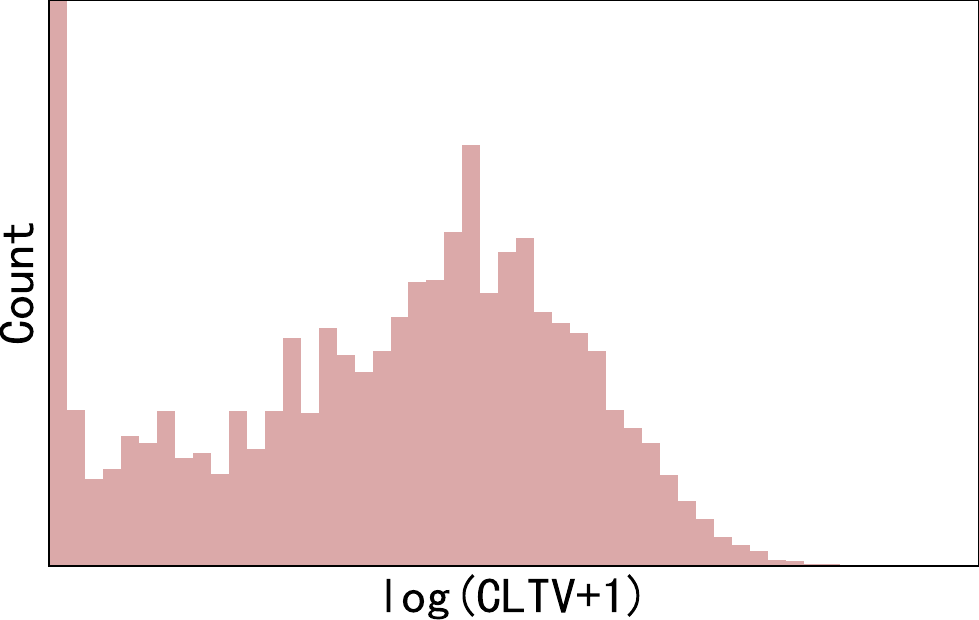}  }     
	\subfloat[Cluster \#2 ] { 
		\label{fig:b} 
		\includegraphics[width=0.48\columnwidth]{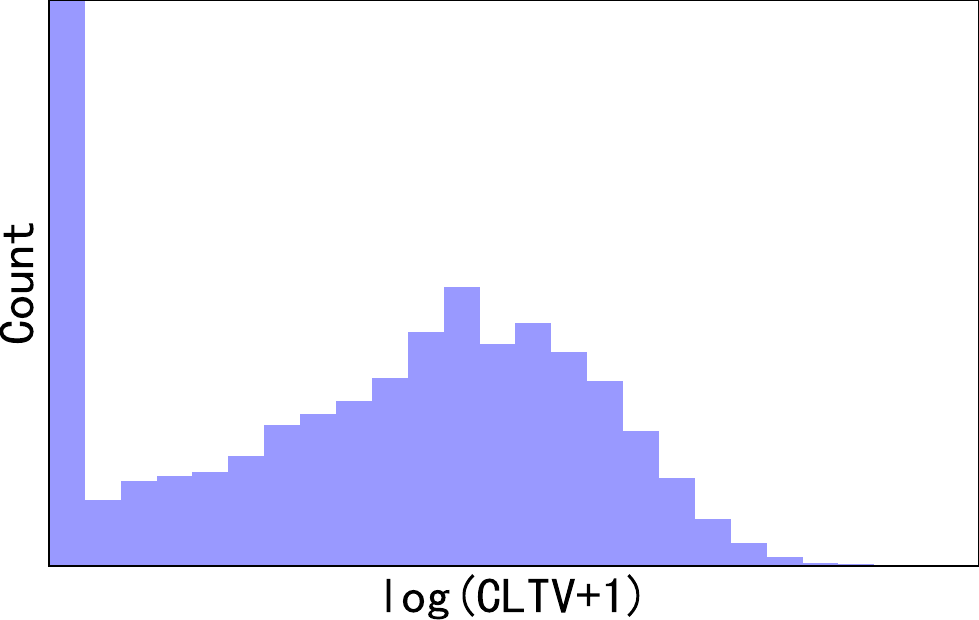}}    
  
 	\subfloat[Cluster \#3] {
		\label{fig:c}
		\includegraphics[width=0.48\columnwidth]{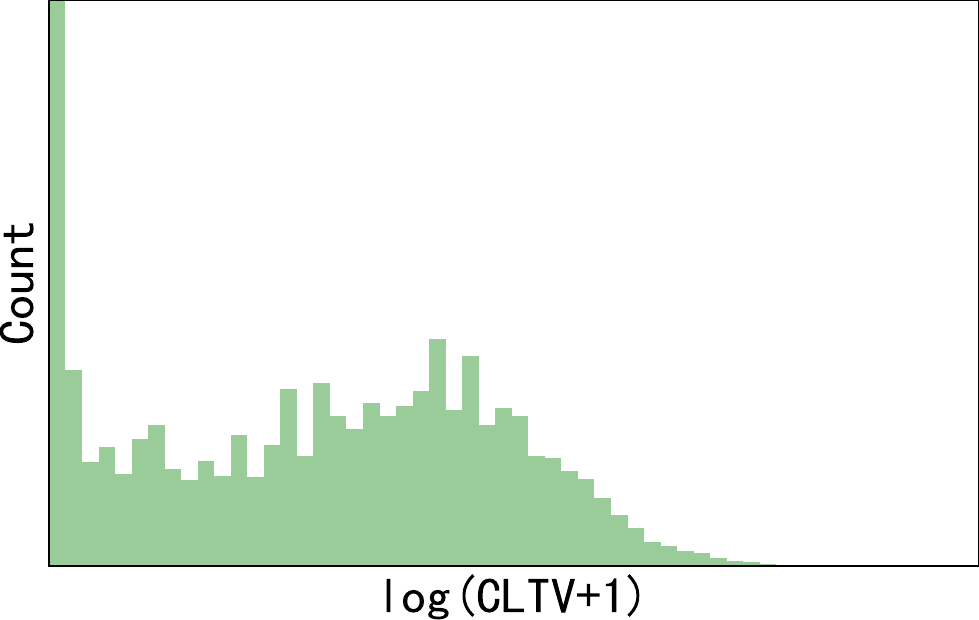}  }     
	\subfloat[Cluster \#4] { 
		\label{fig:d}
		\includegraphics[width=0.48\columnwidth]{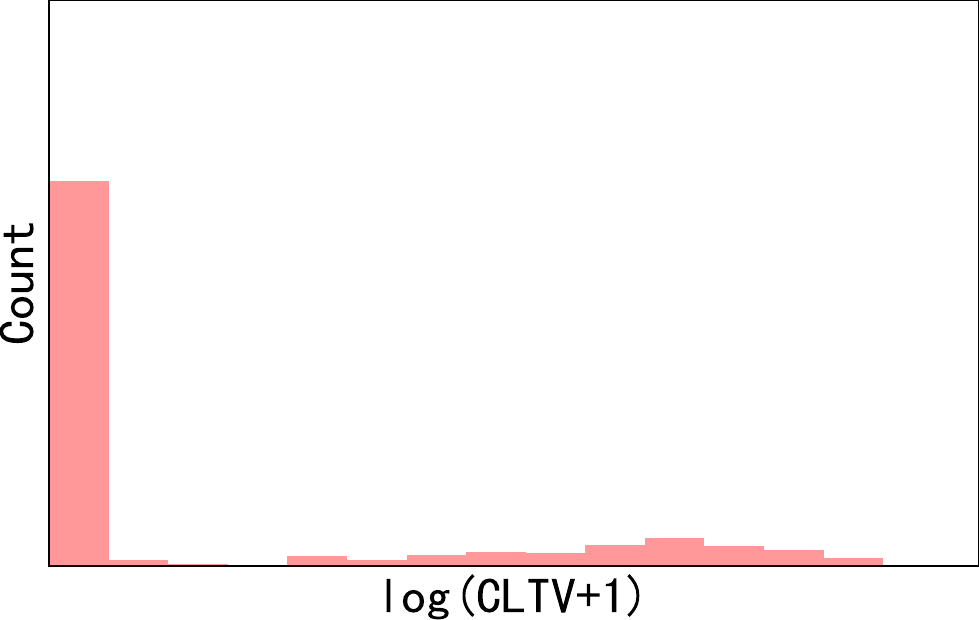}}    
	\caption{The logarithmic CLTV distribution of a large-scale financial platform, with four clusters representing user groups obtained by the clustering algorithm. The x-axis represents $\log(CLTV+1)$, and the y-axis shows the sample count.}  
	\label{distribution}     
\end{figure}

Recently, many efforts have been made to deal with the distribution of CLTV. 
On the whole, these methods can be divided into two categories. 
The first category introduces a deep probabilistic model for CLTV modeling. Zero-inflated lognormal (ZILN)~\cite{wang2019deep} was commonly used to predict CLTV, which employs a deep neural network to model the zero-inflated lognormal probabilistic distribution. With inputting user and item attributes, the deep learning model can predict CLTV for a particular user behavior on a specific item. However, the real-world CLTV distribution is complex, with significant differences in the distribution of different user groups. 
For example, we divide users on a large-scale financial platform into four user clusters based on their attributes and further illustrate the CLTV distribution of each user cluster in Fig.~\ref{distribution}. Based on the notable distribution difference between user groups, we can conclude that utilizing one network to learn the related parameters for all users may lead to insufficient learning. 
Another category is dividing the training samples into several groups according to their continuous CLTV values. Multi Distribution Multi Experts (MDME) model~\cite{li2022billion} divides data examples into multiple predefined sub-distributions based on user CLTV values, and each sub-distribution further contains numerous predefined buckets. The model aims to determine which sub-distribution the user belongs to and the optimal bucket in that sub-distribution.
Nevertheless, simply dividing users into several groups requires rule-based bucketing operations and involves extremely imbalanced classification errors due to numerous zero-value values and high-value users. Moreover, even with the equal-frequency bucketing operation, the CLTV distribution within the buckets might still be uneven, leading to errors in bucket-normalized bias.
Despite existing efforts, an adaptive way to deal with CLTV prediction is still required.

To tackle the above limitations, we introduce a novel framework named \textbf{Opt}imal \textbf{Dist}ribution Selection (OptDist) for CLTV prediction in this paper. Inspired by the intrinsic adaptive performance on different data samples of AutoML~\cite{zhao2021autoloss,optfs,optinter}, OptDist adopts a distribution selection network to automatically select sub-distribution parameters for each example in a differentiable manner. Specifically, instead of using one distribution for all the data examples, we train multiple candidate sub-distribution networks (SDNs) for modeling the CLTV probabilistic distribution in the distribution learning module (DLM) following a divide-and-conquer manner. Notice that each SDN concentrates on training a possible set of probabilistic distribution parameters, thus reducing the complexity of the overall CLTV modeling. Moreover, unlike existing methods that manually partition training examples into different sub-distributions, OptDist introduces a distribution selection module (DSM) that adaptively selects one of the sub-distributions for each individual training example with Gumbel-Softmax~\cite{jang2016categorical} operation. Therefore, at the inference stage, we can use only the optimal sub-distribution selected by the DSM for each predicted instance without creating a gap between training and inference. However, training DSM and the DLM still poses a challenge in this framework due to the different parameter sets in these two modules. We thus propose a novel alignment mechanism to address the issue, which aligns the probability output by DSM to the distribution of loss output by DLM.
The main contributions are summarized as follows:
\begin{itemize}
[topsep=0pt,noitemsep,nolistsep,leftmargin=*]

\item We propose a novel end-to-end CLTV prediction framework named OptDist. Our OptDist explores multiple candidate probabilistic distributions and selects the optimal sub-distribution for each example, which can deal with the complex and mutable distribution of customer lifetime value.  

\item We design two modules, DLM and DSM, respectively, to learn the sub-distribution and distribution selection. We propose an alignment mechanism connected with two modules to train the framework. With two modules and an alignment mechanism, OptDist can adaptively select the optimal sub-distribution for each data example.

\item We conduct comprehensive experiments on two public datasets and a private industrial dataset to verify the superiority of our proposed OptDist model over baselines. Moreover, we have employed OptDist on a large-scale financial platform for marketing campaigns. The online A/B testing results also demonstrate the effectiveness of OptDist.
\end{itemize}

%We organize the rest of the paper as follows. In Section 2, we briefly introduce works related to our work. Section 3 demonstrates the problem formulation first and then discusses the OptDist in detail. In Section 4, extensive experiments are conducted to verify our OptDist. Finally, we conclude this paper in Section 5.

\section{Related work}
In this section, we give a brief review of some related work. Our work is related to two topics: CLTV prediction and AutoML for the recommendation. We summarize the work in the following.

\subsection{CLTV Prediction}

User response modeling is essential for online marketing and recommendation systems. Traditional methods, such as click-through prediction or conversion prediction models, have shown limitations in business scenarios that aim to maximize Gross Merchandise Volume (GMV). Hence, some work~\cite{htlnet} emphasized the necessity to estimate revenue for recommendation systems, accounting for user behavior, conversion rate, and click rate in a ranking model.
To improve ROI and GMV, CLTV estimation emerged as an essential metric to evaluate commercial impact. Conventional statistic-based methods, including RFM~\cite{fader2005rfm} and Pareto/NBD
~\cite{schmittlein1987counting,fader2005counting,bemmaor2012modeling} models, mainly focus on historical data but neglect rich user attribute information. Hence, some work incorporate the user information into the prediction model. Two-stage modeling approaches to predict both the likelihood of purchase and the value of customers are proposed in~\cite{vanderveld2016engagement,drachen2018or}. Word2vec~\cite{mikolov2013efficient} is leveraged for creating user embeddings to predict CLTV~\cite{chamberlain2017customer}. Besides, some work investigated various sequential models for behaviors in CLTV~\cite{bauer2021improved,xing2021learning}, combining RNNs with gradient boosting machines (GBMs)~\cite{bauer2021improved} to capture historical customer behavior, or employing wavelet transforms and attention-based GRU for analyzing user behavior sequences~\cite{xing2021learning}. MarfNet~\cite{yang2023feature} addresses the feature missing problem in the CLTV modeling. These works are perpendicular to our study and can potentially be combined with our method for further improvements.
Introduced a prior distribution, ZILN~\cite{wang2019deep} gives a multi-task solution for CLTV prediction combined classification of returning customers and prediction of returning customer spend. ExpLTV~\cite{outofbox} further extends the ZILN to both game whale detection and CLTV prediction. Moreover, Order Dependency Monotonic Network~\cite{li2022billion} designed the order dependency monotonic network (ODMN) for modeling ordered dependencies between CLTVs to predict the value of CLTV in different periods. At each period, it uses a multi-distribution multi-expert (MDME) module that predicts the classification probabilities with pre-defined buckets and uses them to select proper experts to predict CLTV in certain ranges.

\subsection{AutoML for Recommendation}

In recent years, Automated Machine Learning (AutoML) techniques have gained considerable attention in the recommendation domain for their ability to automatically and efficiently find the best machine learning models and improve performance~\cite{survey,automl,autoopt}. Most previous research focuses on parameter searching such as embedding size~\cite{zhaok2021autoemb,zhao2020memory,optembed} and the process patterns of features including feature bucketing, interaction~\cite{guo2021embedding,luo2019autocross}.
As to embedding size search, AutoEmb~\cite{zhaok2021autoemb} proposed an AutoML-driven approach that decides the optimal embedding size for user/item feature fields based on the contextual information and their popularity. AutoDim~\cite{zhao2020memory} is also proposed for embedding size searching, which learns multiple candidate embeddings with different embedding sizes for each feature field and uses the Gumbel-softmax trick to select the best embedding size.
To achieve automated continuous feature discretization and embedding for enhancing model performance, AutoDis~\cite{guo2021embedding} is proposed by incorporating meta-embedding and automatic discretization modules. AutoCross~\cite{luo2019autocross} focused on the automatic feature interaction operation, which performs beam search in a tree-structured space and generates high-order cross features. Different from the aforementioned studies that focused on feature processing and representation learning, AutoLoss~\cite{zhao2021autoloss} focuses on the search for an appropriate loss function, which involves multiple candidate loss functions and a controller to determine their probabilities. OptDist first introduces AutoML technique to CLTV prediction, which adaptively searches the optimal distribution for data instances.

\begin{figure*}[!htbp]
  \centering
  \includegraphics[width=0.85\linewidth]{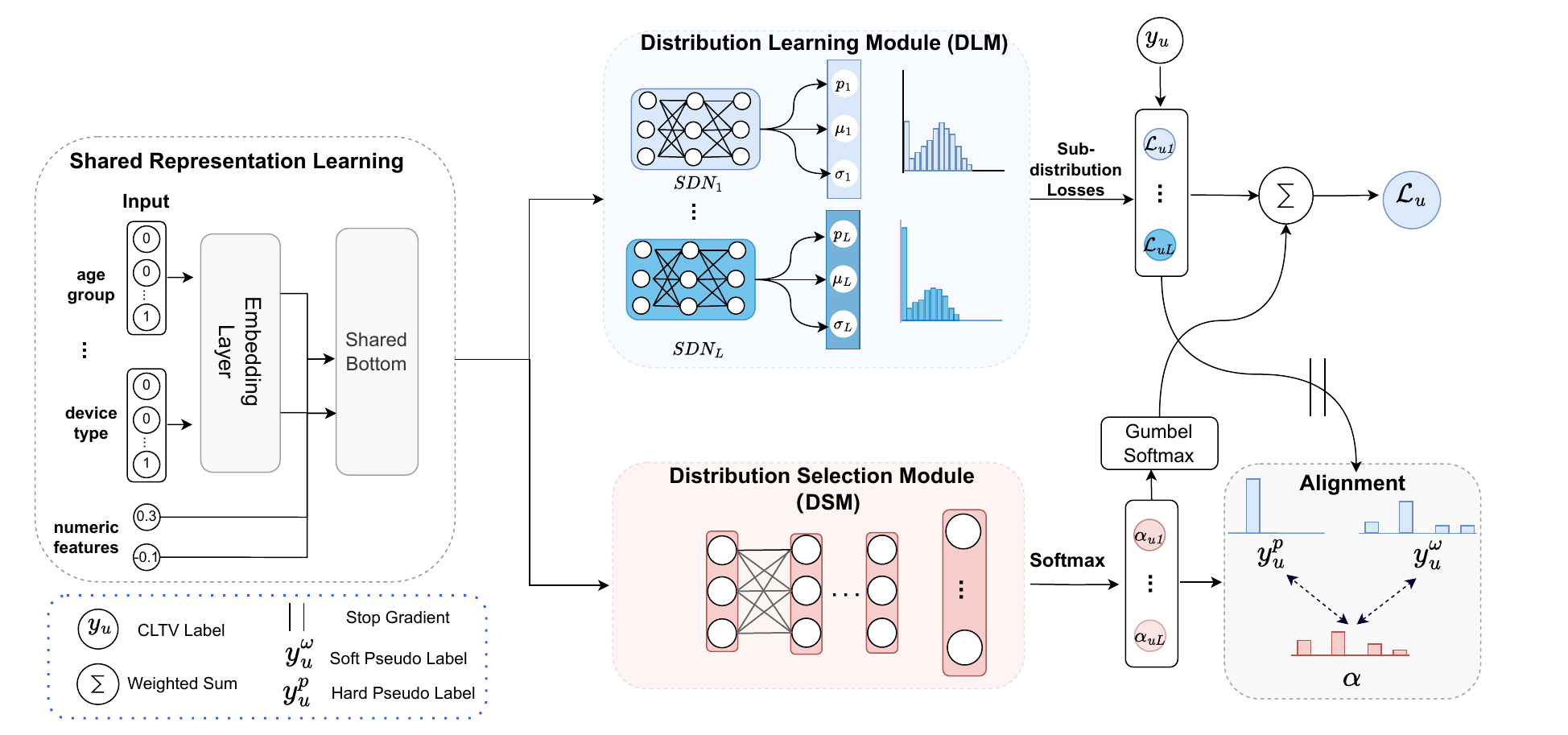}
  \caption{The overall framework of our proposed OptDist.}
  \Description{abt_res}
  \label{OptDist}
\end{figure*}

\section{Method}
\label{sec:method}
The overall framework of OptDist is illustrated in Fig.~\ref{OptDist}. OptDist mainly consists of shared representation learning, a distribution learning module (DLM), and a distribution selection module (DSM). The shared representation learning transforms the original features into dense vectors. DLM comprises multiple sub-networks learning the parameters of one particular probabilistic distribution. The DSM contains a distribution selection network that aims to select an optimal candidate sub-distribution from DLM for each data instance. Then, we describe the alignment mechanism and how to optimize our method.

\subsection{Problem Formulation}
\label{problems}
\subsubsection{Customer Lifetime Value Prediction}
We first give the formulation of CLTV prediction problem. Given a set of $N$ users $\mathcal{U}$ and the total revenue a user $u$ brings to the platform in a fixed time window $d$. Each sample in the training dataset $\mathcal{D} = \{(\mathbf{x}_u, y_u)| u \in \mathcal{U}, y_u \in [0,+\infty) \}$ contains the input feature $\mathbf{x}_u$ and the CLTV label $y_u \ge 0$. In general, we predict the CLTV with the model $f(\cdot)$, which can be formulated as follows,

\begin{gather}
\label{eq:ltv}
    \hat{y}_u = f(\mathbf{x}_u\mid\mathcal{D},{\Theta}), \\
    \mathcal{L}(\hat{\mathbf{y}},\mathbf{y})=\sum_{u=1}^{\mathcal{U}}\mathcal{L}_u(\hat{y}_u,y_u),
\end{gather}
where $\hat{y}_u$ is the prediction CLTV, i.e. pLTV, $\Theta$ denotes the parameters of the model, and $\mathcal{L}$ is loss function for each users $\mathcal{L}_u$.

\subsubsection{Optimal Distribution Selection} Based on Eq.~\ref{eq:ltv}, ${\Theta}$  is usually a probabilistic model that is hard to capture CLTV distribution as discussed above. We thus divided a single probabilistic model into a series of sub-distribution models ${\Theta}=[\theta_1, \cdots, \theta_L]$. We construct the distribution selection as learning a mask vector $\bm{\pi}_u$ for particular user and denotes $\tilde{\Theta} = \bm{\pi}_u \odot \Theta = [\pi_{u,1}\theta_1, \cdots,\pi_{u,L}\theta_L]$. With the notations defined above, our optimal distribution selection problem is formally defined as follows:

\begin{gather}
\label{eq:def}
\theta^\star = \argmin_{\bm{\pi}_u,\Theta}\mathcal{L}_u(f(\mathbf{x}_u \mid \mathcal{D}, \bm{\pi}_u\odot\Theta),y), \\
s.t. \quad \sum_{l=1}^L \pi_{u,l} = 1, \quad \pi_{u,l} \in \{0,1\} \notag,
\end{gather}
where $L$ is the number of candidate sub-distributions to select and $\theta_l$s are initialized with different seeds to increase diversity between sub-distributions.

\subsection{Shared Representation Learning}
The input feature $\mathbf{x}_u \in \mathcal{R}^m$ mainly consists of categorical and continuous features. Usually, each categorical feature $x_i$, such as the user's city or gender, will be embedded into a low dimensional vector $\mathbf{e}_i$ via the embedding table. For continuous features such as the number of visits, we normalize them using Z-score~\cite{al2006data} and treat them as a one-dimensional vector $\mathbf{e}_j$. All feature vectors are concatenated together:
\begin{equation}
    \textbf{h} = [\mathbf{e_1}, \mathbf{e_2},...,\mathbf{e_m}].
    \label{concate}
\end{equation}

Subsequently, $\textbf{h}$ will be transformed by shared bottom layers to generate a shared sample representation and fed into the DLM and DSM, respectively. Note that $\textbf{h}$ is input for DSM and DLM. Therefore, other representation learning modules for specific tasks can be easily plugged into our framework, such as MarfNet~\cite{yang2023feature} for missing feature problems, CDAF~\cite{cdaf} for cross-domain adaption, and so on.

\subsection{Distributions Learning Module}

As shown in Fig.~\ref{distribution}, capturing the CLTV distribution within one distribution is challenging due to the large fraction of zero-consumption customers and a small number of high-value customers for most business scenarios. Therefore, we adopt the idea of \textit{Divide-and-Conquer}, which introduces several neural networks to learn part of the distribution. 

Specifically, we assume that the overall complex distribution of CLTV comprises several sub-distributions, and each user belongs to one of these sub-distributions. We use several sub-networks, denoted as sub-distribution networks (SDNs), to model each sub-distribution.  Each SDN focuses on learning from a subset of users with similar distributions, thus avoiding the impact of significant distribution differences between users on the effectiveness of parameter learning. Therefore,  OptDist reduces the difficulty of overall probabilistic distribution modeling. Notice that different from the method in~\cite{li2022billion}, which manually divides samples into sub-distributions based on the ground truth in advance, OptDist searches the optimal sub-distribution to which each user belongs automatically.

Note that there are two critical issues in this module. First, it is crucial to determine the sub-distribution network, which indicates how to model the distribution. As is previously stated, zero-inflated lognormal distribution~\cite{wang2019deep} is specially proposed for the CLTV distribution. The ZILN loss alleviates the problem of commonly used MSE loss being overly sensitive to extreme values. Therefore, we adopt the network for ZILN loss in our framework as SDNs. Second, how many neural networks will be set can decide the search space in our OptDist. In the training, each user representation $\mathbf{h}_u$ will be fed into the fixed number of candidate SDNs, obtaining a set of different ZLIN distribution parameters $\{\theta_1 = (p_1,\mu_1,\sigma_1), \theta_2 = (p_2,\mu_2,\sigma_2),...,\theta_L = (p_L,\mu_L,\sigma_L)\}$. As a result, the search space can be $N^L$, which poses a significant challenge to search in a large-scale platform. Hence, there is a trade-off to determine the $L$. 
If $L$ is too large, it will increase the burden of searching, while too small will lead to the model's inability to fit complex distributions. Thus, we set $L$ as a hyperparameter, which can be efficiently explored in our framework. In conclusion, we calculate the negative log-likelihood loss of user $u$ for each SDN$_i$:
\begin{equation}
     {\mathcal{L}_{u,i}} = \left\{
	\begin{aligned}
		& -\log({p_{u,i}}) + \log({y_u \sqrt{2\pi}\sigma_{u,i}  })+{\frac{(\log y_u - \mu_{u,i})^2}{2{\sigma_{u,i}}^2}}  , &  C_u = 1 \\
		& -\log(1 - p_{u,i})  , & C_u = 0
	\end{aligned}
	\right.
\label{ziln}
\end{equation}
where $C_u$ denotes whether the user $u$ is converted and $y_u$ is the CLTV label. Then, we re-write Eq.~\ref{eq:def} to obtain the specific loss function $\mathcal{L}_{u}$ by calculating the weighted sum of losses for each SDN for each user:
\begin{equation}
\mathcal{L}_{u} = \sum_{i=1}^L {\pi_{u,i} \cdot \mathcal{L}_{u,i}},
\label{loss_s}
\end{equation}
where $\pi_{u,i}$ is weight of user $u$ for the $i$-th candidate sub-distribution. $\pi_{u,i}$ is obtained from the output of the DSM module, which is discussed in the next section.

\subsection{Distribution Selection Module}
\label{dsm_section}

Tackling the optimal distribution selection problem in Eq.~\ref{eq:def} is challenging in our OptDist. As a potential solution, the reinforcement learning agent can only receive the reward until the optimal distribution network is selected. This prevents the direct application of reinforcement learning due to delayed rewards. Hence, to determine the weights in Eq.~\ref{loss_s}, OptDist introduces a distribution selection module following the design in~\cite{zhao2021autoloss}. 

Our OptDist adopts the Multi-Layer Perceptron (MLP) as the optimal distribution selection network. The output of the optimal distribution selection network is formulated as follows:
\begin{equation}
\bm{\alpha}_u = softmax(\mbox{MLP}(\mathbf{h}_u|\theta_{mlp})),
\label{softmax_formula}
\end{equation}
where $\theta_{mlp}$ is the parameters of MLP. It is worth noting that the selection network can be easily substituted with other more powerful models~\cite{guo2017deepfm,wang2021dcn}, which is out of the scope of this paper.

However, using softmax operations might produce relatively smooth weights. This may lead to the selected SDN training being influenced by the losses of other SDNs, resulting in sub-optimal results. Gumbel-max sampling~\cite{gumbel1954statistical} is a technique that enables hard selection:

\begin{gather}
S_{u} = \mbox{one\_hot}(\argmax_{i}[\log{\alpha_{u,i}} + g_{u,i}]), \\
g_{u,i} = -\log(-\log(U_{u,i})) \notag, \\
U_{u,i} \sim Uniform(0,1) \notag.
\end{gather}
However, this discrete selection is non-differentiable due to the $\argmax$ operation. To tackle this, we employ the straight-through Gumbel-softmax~\cite{jang2016categorical}:

\begin{equation}
	\pi_{u,i} = \frac{\exp((log(\alpha_{u,i}) + g_{u,i})/\tau)}{\sum_i {\exp((log(\alpha_{u,i}) + g_{u,i}})/\tau)},
 \label{gumbel_softmax}
\end{equation}
where $\tau$ is the temperature parameter, which controls the approximation degree between the Gumbel-softmax distribution and the discrete distribution. As $\tau$ approaches 0, the effect of Eq.~\ref{gumbel_softmax} becomes closer to the $\argmax$ operation, thus getting the mask vector $\bm{\pi}_u$.

\subsection{Alignment Mechanism}

As in Eq.~\ref{loss_s}, we need to optimize the loss function with outputs from both modules. Specifically, the DLM module will update distribution network parameters, while DSM also updates the selection policy accordingly, making optimization difficult and sub-optimal. To further enhance the optimization, we propose an alignment mechanism inspired by meta pseudo labels~\cite{meta}.

\begin{figure}[!t]
  \centering
  \includegraphics[width=0.84\linewidth]{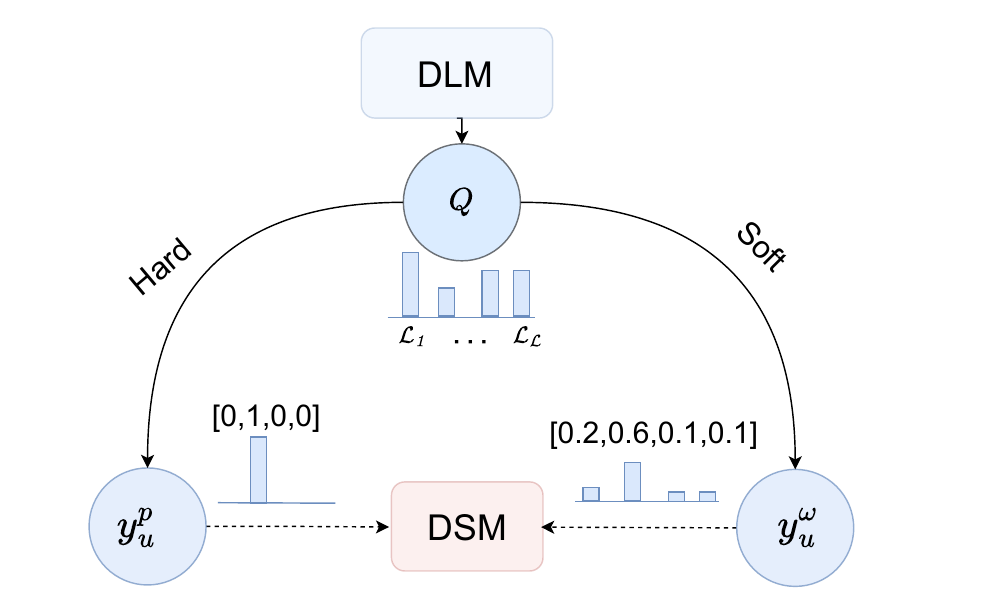}
  \caption{The alignment mechanism between DSM and DLM. $Q$ denotes the set of SDN's losses. }
  \Description{alignment}
  \label{alignment_loss}
\end{figure}

In our OptDist, each SDN within the DLM module focuses on training with instances allocated to that sub-distribution by the DSM. However, two individual sets of parameters in our OptDist interfere with each other during optimization.
Meanwhile, there is a lack of explicit supervised signals for the DSM, which is hard to align with the output of DLM. It is challenging to train both DSM and DLM well merely relying on the loss $\mathcal{L}_u$. By normalizing the loss values generated by different SDNs for each user, the alignment mechanism can generate pseudo labels to guide the training of the DSM, reducing the difficulty of searching for the optimal sub-distribution for DSM. As Fig.~\ref{alignment_loss} illustrated, when a set of loss values on possible distribution $\textbf{Q} = \{\mathcal{L}_{u,i}\}_{i=1}^{L}$ is given, we can obtain the hard pseudo labels $\mathbf{y}_u^{p}$ from these loss values:

\begin{equation}
\mathbf{y}_u^{p} = \mbox{one\_hot}({\argmin}_{i} ({{\mathcal{L}}_{u,i})}). 
\label{hard_label}
\end{equation}

First, the hard label $\mathbf{y}^p$ can construct a cross-entropy loss. In addition, considering that in practical applications, the classification of CLTV is imbalanced, which may result in the cross-entropy of high-value users being overlooked, making it challenging for DSM to distinguish them. Therefore, to mitigate this issue, we have introduced a variant of focal weight~\cite{lin2017focal} in the cross-entropy loss. The loss function can be defined as:

\begin{equation}
\mathcal{L}_u^{CE}=\sum_{i=1}^L -y_{u,i}^p(1-\alpha_{u,i})^2\log(\alpha_{u,i}).
\end{equation}

Then, we generate soft labels based on the losses for each sub-distribution:
\begin{gather}
y_{u}^{\omega} = softmax(-\mathcal{L}_u) = [\omega_{u,1}, \omega_{u,2} ,..., \omega_{u,L} ] \\
\omega_{u,i} = \frac{\exp(-{\mathcal{L}_{u,i}})}{\sum_{j}  \exp(-{\mathcal{L}_{u,j}})}.
\label{soft_label}
\end{gather}
The larger the $\omega_{u,i}$, the more suitable the $i$-th sub-distribution is for user $u$ according to DLM. Then, we adopt Kullback-Leibler (KL) divergence~\cite{PRML} between DLM and DSM:
\begin{equation}
\mathcal{L}_u^{KL}=\sum_{i=1}^{L}\omega_{u,i}\log(\frac{\omega_{u,i}}{\alpha_{u,i}}).
\end{equation}
The advantage of considering both hard and soft labels here lies in that a hard label can make DSM focus on DLM information while ignoring other label information, which is complemented by soft labels.
In summary, the overall loss for OptDist is defined as:

\begin{equation}
    \mathcal{L}^{OptDist} = \frac{1}{N}\sum_{\mathbf{u} \in \mathcal{U}}( \mathcal{L}_{u} + \mathcal{L}_{u}^{CE} + \mathcal{L}_{u}^{KL}  ).
    \label{overall_loss}
\end{equation}

\subsection{Optimization and Inference}

\subsubsection{Optimization Method}

In OptDist, the trainable parameters come from DLM and DSM. We denote the parameters of DLM and DSM as $\Theta_{L}=\{\theta_1,\cdots,\theta_L\}$ and $\Theta_{S}=\{\theta_{mlp}\}$, respectively. Note that $\bm{\pi}_u$ in Eq.~\ref{eq:def} is directly generated by DSM as in Eq.~\ref{gumbel_softmax}. Here, we mainly discuss how to optimize the framework parameters. We form a bi-level optimization problem for our OptDist as follows:

\begin{equation}
\begin{split}
\label{eq:bilevel}
& \min_{\Theta_S}\mathcal{L}^{OptDist}_{val}(\Theta_L^\star,\Theta_S), \\
& s.t.\quad \Theta_L^\star=\argmin_{\Theta_L}\mathcal{L}^{OptDist}_{train}(\Theta_L,\Theta_S^\star),
\end{split}
\end{equation}
where DLM parameters $\Theta_L$ and DSM parameters $\Theta_S$ are considered as the upper- and lower-level variables. However, this formulation increases the complexity of model training. Therefore, We adopt an approximation scheme strategy by differentiable architecture search (DARTS)~\cite{liu2018darts}. All the parameters of OptDist, denoted as $\Theta=\{\Theta_L,\Theta_S\}$, are updated as follows within each mini-batch:

\begin{equation}
    \hat{\Theta} =  \Theta - \eta \cdot \bigtriangledown{\mathcal{L}_{train}^{OptDist}},
    \label{update}
\end{equation}
where $\eta$ is the learning rate. Note that $\Theta_L$ and $\Theta_S$ have shared and independent parameters, where the alignment mechanism alleviates the difficulty of approximating one-level optimization.

\subsubsection{Inference Stage}

For each instance to be predicted $\mathbf{x}$, OptDist first obtains the representation $\mathbf{h}$ through the shared embedding bottom. Then, the representation is fed into DSM, which will output the probability $\alpha$ that this instance belongs to each sub-distribution. Only the optimal SDN's output will be employed for the predicted CLTV calculation in this stage, and the index of that SDN could be obtained with argument max operation on $\alpha$:
\begin{equation}
s = \argmax{([\alpha_1, \alpha_2,...,\alpha_L])}.
\end{equation}
With the index of selected distribution, the model can fetch the optimal distribution parameters $\theta_s = (p_s, \mu_s, \sigma_s) $ from the corresponding SDN's output and combine them with the expectation formula of the log-normal distribution to obtain the estimated CLTV:
\begin{equation}
    \hat{y} = p_s \times \exp(\mu_s + \sigma_s ^2 /2).
\end{equation}

\begin{algorithm}[!t]
	\caption{An Optimization Algorithm for OptDist  }\label{algorithm}
	\KwIn{ Input features $\mathbf{X} = \{\mathbf{x_u}| \forall u \in \mathcal{U}\}$; \\
         number of sub-distribution $L$;\\
         temperature $\tau$;\\

         Initial model parameters $\Theta$;\\
         
         Learning rate $\eta$;}
         
	\KwOut{Predicted CLTV $\hat{y}$\\
            Trained model parameters $\hat{\Theta}$;
            }
 
        $ \mathbf{H} = Emb(\mathbf{X}) $\;
        
	\For{i = 1 to $L$}{
                $(p_i,\mu_i,\sigma_i) \leftarrow SDN_i(\mathbf{H})$\ ;
        }
        $\alpha$ = softmax( MLP($\mathbf{H}$))\;
        
        $\pi = gumbel\_softmax(log(\alpha),\tau)$

        \For{i = 1 to $L$}{
            calculate the negative log-likelihood loss $\mathcal{L}_i$ for the corresponding sub-distribution $SDN_i$ ;
        }
        Generate the hard label and soft label according to Eq. ~\ref{hard_label} and Eq. ~\ref{soft_label} respectively; \ 
        
        Calculate the loss $\mathcal{L}$ according to Eq.~ ~\ref{loss_s} - Eq. ~\ref{overall_loss}.\
        
        $\hat{\Theta}$ $\leftarrow$ $\Theta$ - $\eta$ $\cdot$ $\bigtriangledown{\mathcal{L}}^{OptDist} $\ 
        
        $s = \argmax(\alpha)$ \
        
        $\hat{y} = p_s \times \exp(\mu_s + \sigma_s ^2 /2)$ \

\end{algorithm}

The Algorithm.~\ref{algorithm} summarizes the optimization and inference process of OptDist. In lines 2-4, the sample representation is fed into several sub-networks to obtain the sub-distribution parameters. Since each sub-distribution only needs to handle the sub-problems of modeling the CLTV probability for a part of similar samples, the sub-network can be processed using a relatively simple MLP and can be parallelized, thus not increasing the time complexity of the model. Lines 5 and 6 present the output of DSM, while distribution evaluation is conducted from Line 7 to Line 11. Line 12 carries out the parameter update for the model. Finally, the predicted results are obtained according to Lines 13-14.

\section{experiment}
In this section, we conduct both offline and online experiments to demonstrate the effectiveness of our proposed Optdist and answer the following research questions:
\begin{itemize}
[topsep=0pt,noitemsep,nolistsep,leftmargin=*]
\item \textbf{RQ1}: How does the offline and online performance of our proposed OptDist compare with mainstream baselines?

\item \textbf{RQ2}: How do the key hyper-parameters influence the performance of OptDist? 

\item \textbf{RQ3}:  What is the impact of the alignment mechanism in OptDist on the final result?

\item \textbf{RQ4}: Can OptDist learn optimal sub-distribution?

\end{itemize}

\subsection{Experimental Setup}
\subsubsection{Dataset.} We conduct experiments on two public datasets and one private industrial dataset. In the following dataset, we randomly split them into 7:1:2 as the training, validation, and test sets, respectively.

\textbf{Criteo-SSC.}\footnote{https://ailab.criteo.com/criteo-sponsored-search-conversion-log-dataset/} The Criteo Sponsored Search Conversion (Criteo-SSC) Dataset is a large-scale public dataset, which contains logs obtained from Criteo Predictive Search (CPS). Each row in the dataset represents a user's click behavior on a product advertisement and contains information about the product's attributes and the user's characteristics. The label is whether the click led to a conversion and the corresponding revenue within 30 days. Note that we remove the product price from the features.
    
\textbf{Kaggle.}\footnote{https://www.kaggle.com/c/acquire-valued-shoppers-challenge} The Kaggle's Acquire Valued Shoppers Challenge dataset, hereafter referred to as the Kaggle Dataset, contains transaction records of over 300,000 shoppers at about 3,300 companies. Similar to the experimental setting of the previous research~\cite{wang2019deep}, the task we consider is to predict the total value of a company's products purchased by a user in the year following their initial purchase and focus on the customers whose initial purchase occurred between 2012-03-01 and 2012-07-01. We retain the three companies with the most transactions.

\textbf{Industrial.} The private dataset is collected from a large-scale financial platform that offers mutual funds from various fund companies. Since customers can freely determine their investment amounts, the distribution of their customer lifetime value (CLTV) is very complex.  The dataset consists of over 4.5 million samples, each corresponding to the profiles and access behavior features of a new user who has never invested in the platform. The model aims to predict whether these users will convert within the next 30 days and estimate their corresponding CLTV.

Table.~\ref{tab:commands} summarizes the details of these three datasets.

\begin{table}[!t]
	\centering
	\caption{Dataset Statistics}
	\label{tab:commands}
	\begin{tabular}{cccc}
		\toprule
		   Dataset & Samples & Positive Samples & Positive Ratio    \\

		\midrule
		Criteo-SSC & 15,995,633 & 1,150,996 &  7.20\%  \\
		\midrule
		Kaggle & 805,753 & 726,180 &  90.12\%  \\
		\midrule
		Industrial & 4.535,675 & 287,934 &  6.35\%  \\
		\bottomrule
	\end{tabular}
\end{table}

\subsubsection{Metrics.} The CLTV prediction is continuous, and thus, we use \textbf{MAE} to measure the deviation between the predicted value and the actual CLTV of the user, which is widely used as a metric in regression tasks~\cite{li2023dynamic}. In practical business applications, marketing resources tend to be allocated toward customers with higher CLTV, and thus, the accuracy of ranking users based on the predicted CLTV is more concerned~\cite{yang2023feature}. Following the previous research~\cite{wang2019deep,yang2023feature}, we adopt both \textbf{Spearman rank correlation (Spearman's $\rho$)} and the \textbf{normalized Gini coefficient (Norm-GINI)} to evaluate. Notice that the larger this value, the better the CLTV prediction is. Apart from using these two ranking metrics to evaluate the overall ranking performance of the models on all samples, we also perform evaluations on positive samples separately to compare the distinguishing ability of models for non-zero CLTV samples, which are denoted as \textbf{Norm-GINI(+)} and \textbf{Spearman's $\rho(+)$}.

\begin{table*}[htbp]
  \caption{The overall performance of different models on all datasets. $\uparrow$ indicates that the higher value of the metric is better, while $\downarrow$ signifies the opposite. $\dagger$ indicates statistically significant improvement over the best baseline(p-value < 0.05). }
\begin{tabular}{|c|c|c|c|c|c|c|}
  \hline % horizontal line
  Dataset & Model & MAE $\downarrow$ & Norm-GINI $\uparrow$  & Spearman's $\rho$ $\uparrow$  & Norm-GINI(+) $\uparrow$  & Spearman's $\rho$(+) $\uparrow$\\ %first cell is occupied by the multirow
  \hline
  \hline
   \multirow{5}{*}{\centering Criteo-SSC} & Two-stage & 21.719  & 0.5278  & 0.2386  & 0.2204  & 0.2565 \\ %end line
  \cline{2-7} %short partial horizontal lines from column 2 to column 5
   & MTL-MSE & 21.190  & 0.6330  & 0.2478  & 0.4340  & 0.3663  \\ %first cell is occupied by the multirow
  \cline{2-7} 
  & ZILN & 20.880 & 0.6338  & 0.2434  & 0.4426  & 0.3874 \\
    \cline{2-7} 
  &MDME &  16.598 & 0.4383  & 0.2269  & 0.2297  & 0.2952 \\

  \cline{2-7}

    &MDAN &  20.030 & 0.6209  & 0.2470  & 0.4128  & 0.3521 \\

  \cline{2-7}
  
  & \textbf{OptDist} &  $\textbf{15.784 }^\dagger$ & $\textbf{0.6437 } ^\dagger$ & \textbf{0.2505 } & \textbf{0.4428} & \textbf{0.3903 }\\
  \hline
    \hline

  \multirow{5}{*}{\centering Kaggle} & Two-stage & 74.782 & 0.5498 & 0.4313  & 0.5505  & 0.4596 \\ %end line
  \cline{2-7} %short partial horizontal lines from column 2 to column 5
   & MTL-MSE & 74.065  &  0.5503  & 0.4329  &   0.5349 & 0.4328 \\ %first cell is occupied by the multirow
  \cline{2-7} 
  & ZILN & 72.528 & 0.6693  & 0.5239  & 0.6627  & 0.5303 \\
    \cline{2-7} 
  & MDME &  72.900 &  0.6305 & 0.5163 & 0.6213 &  0.5289 \\
  \cline{2-7}

   & MDAN &  73.940 &  0.6648 & 0.4367 & 0.6680 &  0.4567 \\
  \cline{2-7}
  & \textbf{OptDist} &  $\textbf{70.929}^{\dagger}$ & $\textbf{0.6814 }^{\dagger}$ & \textbf{0.5249}  & $\textbf{0.6715}^{\dagger}$  & $\textbf{0.5346}^{\dagger}$ \\
  \hline
    \hline

  \multirow{5}{*}{\centering Industrial} & Two-stage & 0.887 & 0.6670 & 0.0781  & 0.5588  & 0.4467 \\ %end line
  \cline{2-7} %short partial horizontal lines from column 2 to column 5
   & MTL-MSE & 0.548  &  0.7194  & 0.1161  &   0.5575 & 0.4274 \\ %first cell is occupied by the multirow
  \cline{2-7} 
  & ZILN & 0.389 & 0.7854  & 0.1208  & 0.5899  & 0.5401 \\
    \cline{2-7}

  &MDME &  0.419 & 0.7277  &  0.1229  & 0.5609 & 0.5119 \\
  \cline{2-7}
  &MDAN &  0.437 & 0.7629  &  0.1214  & 0.5816 & 0.5383 \\
  \cline{2-7}
  & \textbf{OptDist} &  $\textbf{0.322}^{\dagger}$ & $\textbf{0.8283 }^{\dagger}$ & $\textbf{0.1282}^{\dagger}$   & $\textbf{0.6271}^{\dagger}$  & $\textbf{0.5476}^{\dagger}$ \\
  
  \hline
\end{tabular}
\label{evaluation}
\end{table*}

\subsubsection{Baselines.} We compared our proposed OptDist with several state-of-the-art CLTV prediction approaches. Note that some approaches focusing on representation learning~\cite{xing2021learning,chamberlain2017customer,yang2023feature} are not included here. The baselines are summarized as follows: 
\begin{itemize}
[topsep=0pt,noitemsep,nolistsep,leftmargin=*]
    \item Two-stage~\cite{drachen2018or}. It decomposes the CLTV prediction into two tasks: the first task is a classification task predicting whether a user will churn or not, and the second task is a regression task predicting the revenue that the user brings.
    \item MTL-MSE~\cite{ma2018entire}. It estimates conversion rate and CLTV with MSE loss according to the multi-task learning paradigm.
    \item ZILN~\cite{wang2019deep}. ZILN assumes that the long-tailed CLTV distribution follows a zero-inflated log-normal distribution and uses a DNN to estimate the mean $\mu$, standard deviation $\sigma$, and conversion rate $p$ for the samples.

    \item MDME~\cite{li2022billion}. This baseline divides the training samples by CLTV into multiple sub-distributions and buckets, and constructs corresponding classification problems to predict the bucket a sample belongs to. In the next stage, the bias within the bucket is estimated so that the samples obtain a fine-grained CLTV value.

    \item MDAN~\cite{MDAN}. MDAN predicts predefined LTV bucket labels using a multi-classification network and leverages a multi-channel learning network to derive embeddings for each bucket. The final sample representation is obtained by fusing these embeddings with the classification network's output through a weighted sum, which is then utilized for CLTV prediction.

\end{itemize}

\subsubsection{Implementation Details.}
In this subsection, we provide the implementation details. For a fair comparison, in all experiments of OptDist and all baselines, the learning rate was chosen from $[\text{5e-4}, \text{1e-3}, \text{1.5e-3}, \text{2e-3}, \text{2.5e-3}]$. For both the two public datasets, the batch size was set to 2048, and the embedding size was 5. For the industrial dataset, the batch size was set to 512 and the embedding size was 12. For ZILN, MSE, and MTL-MSE, the size of the MLP part was set to $[64, 64, 64]$ for the Kaggle dataset, $[512, 256, 64]$ for the Criteo-SSC dataset, and $[512, 256, 128]$ for the industrial dataset. For OptDist's each SDN, MDME's each bucket network, and MDAN's each channel network, the corresponding size was set to $[64, 32, 32]$, $[256, 128, 64]$, and $[256, 128, 64]$, respectively.

Our implementation is based on Tensorflow \cite{tensorflow} and all experiments are conducted on a Linux server with one Nvidia-Tesla V100-PCIe-32GB GPU, 128GB main memory, and 8 Intel(R) Xeon(R) Gold 6140 CPU cores. 

%For ZILN\footnote{https://github.com/google/lifetime\_value} and MTL-MSE\footnote{https://github.com/shenweichen/DeepCTR}, we reuse the official or common implementation. Due to the lack of available implementation for the Two-Stage, MDME and MDAN, we re-implement them based on the details provided by the authors.

\subsection{Performance Comparison(RQ1)}

In Table~\ref{evaluation}, we present the evaluation results of each model on testing sets of all datasets, respectively. Based on these results, we have the following insightful observations: 
\begin{itemize}
[topsep=0pt,noitemsep,nolistsep,leftmargin=*]
    \item  MTL-MSE has better performance in the overall dataset evaluation compared to the two-stage model.  In evaluating the positive sample space, MTL-MSE is not necessarily superior to the two-stage models. This is because, in the two-stage methods, the learning of CLTV is more sufficient for those users with a high predicted conversion rate, and the performance of MTL-MSE might be affected by the seesaw phenomenon. 
    \item  The overall performance of ZILN is better than that of both two-stage and MTL-MSE, indicating that modeling the probability distribution of CLTV can alleviate the problem of MSE being sensitive to extreme values.
    \item The performance of MDME is unstable across the datasets. For example, it has a small MAE on the Criteo-SSC dataset, but both the Norm-GINI and Spearman's rank correlation metrics are poor, indicating that its ranking ability is weak on this dataset. On the Industrial dataset, although the overall Spearman's $\rho$  is relatively better than ZILN, the overall Norm-GINI as well as the ranking metrics in the positive sample space are significantly weaker than those of ZILN and OptDist. This is because MDME needs to predict the sub-distribution and bucket to which the samples belong, as well as the position of the samples within the bucket, which might lead to the amplification of accumulated errors.  Although MDAN achieves more stable results compared to MDME by fusing the embeddings of multiple channels, it still requires predefined bucketing. Additionally, it uses a single prediction network to estimate CLTV after integrating representations from various channels, which may not effectively capture complex CLTV distributions, limiting the model's performance.
    \item  Our proposed OptDist outperforms baselines across the three datasets. This indicates that by adaptively learning different sub-distribution parameters and selecting the optimal sub-distribution, it is possible to decompose the complex overall distribution into multiple relatively easy-to-learn subproblems, thereby improving the model's predictive performance. Moreover, OptDist does not require additional predefined bucketing of samples, which enables incremental training and quick deployment to other new scenarios. Due to the typically high proportion of zero-value samples in CLTV estimation problems, achieving equal frequency bucketing for MDME is difficult. Moreover, even if the positive sample bucketing is at equal frequency, the CLTV distribution within the bucket may not be uniform, leading to inaccurate bucket bias estimation. We argue that pre-defined bucketing and the estimation method using normalized positions within buckets may not be entirely suitable in this situation.
    
\end{itemize}

\subsection{Hyper-Parameter Sensitivity Analysis(RQ2)}

In the DSM of our OptDist, Gumbel-softmax's temperature coefficient affects each SDN's weights in $\mathcal{L}_{u}$. Moreover, the number of sub-distributions in DLM is also a critical hyper-parameter. Therefore, this section investigates how these two hyper-parameters affect our framework. Note that we mainly focus on the Norm-GINI evaluation of the overall samples in practical business scenarios, as it indicates whether the model can help allocate marketing resources to users with the highest CLTV~\cite{yang2023feature}. Therefore, concerning this metric, we mainly discuss the influence of different parameters on OptDist performance.
\begin{figure}[t]
  \centering
  \subfloat[Number of Sub-distributions]{
    \includegraphics[width=0.48\columnwidth]{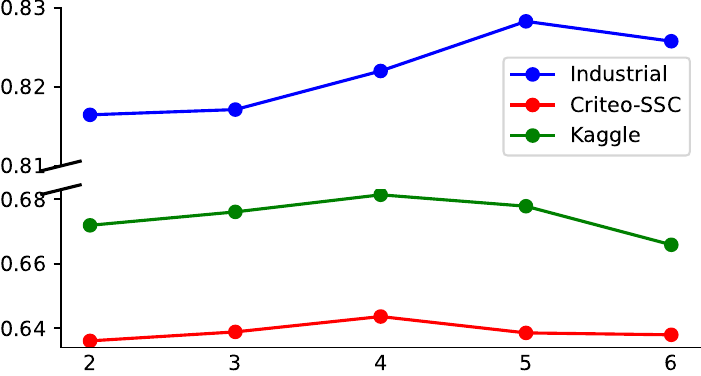}

  }
  \subfloat[Gumbel-softmax temperature]{
    \includegraphics[width=0.48\columnwidth]{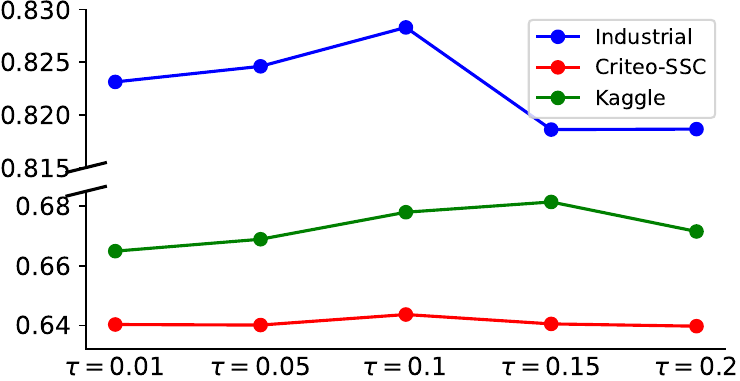}

  }
  \caption{Norm-GINI of OptDist with different hyper-parameters on the three datasets.}
  \Description{abt_res}
  \label{parameter}
\end{figure}
For each dataset, we vary the number of sub-distributions in the set \{2, 3, 4, 5, 6\}. 
In general, on the one hand, a dataset with a more complex overall distribution may contain more sub-distributions, requiring more SDNs for modeling. On the other hand, an increase in the number of sub-distributions also increases the difficulty of learning with DSM. 
In Fig.~\ref{parameter} (a), we display the ranking performance of the model with different sub-distribution settings. For the Kaggle and Criteo-SSC datasets, the performance of the model is best when the number of sub-distributions is 4, while for the industrial dataset, the performance is best when the number of sub-distributions is 5.  Fig.~\ref{parameter}(b) shows the performance of the framework under different Gumbel-softmax temperature coefficients. When the temperature coefficient is too high, the sampling probability becomes smoother, and the sample cannot focus on the training of the sub-distribution it selected, thus affecting the performance.

\begin{table}[!t]
	\centering
	\caption{Norm-GINI of OptDist and its derivations on different datasets.}
	\label{res_2}
	\label{ablation_results}
	\begin{tabular}{cccc}
		\toprule
		   Method & Criteo-SSC & Kaggle & Industrial  \\

		\midrule
		OptDist & \textbf{0.6437} & \textbf{0.6814} & \textbf{0.8283}  \\
		\midrule
		  (w/o) Gumbel-softmax & 0.6389 & 0.6628 & 0.8231  \\

		\midrule
		(w/o) $\mathcal{L}^{KL}$ & 0.6382 & 0.6786 &  0.8158  \\

		\midrule
		(w/o) $\mathcal{L}_u^{CE}$ &  0.6366 & 0.6761 &  0.8107  \\
  		\midrule
		(w/o) $\mathcal{L}^{KL}$ + $\mathcal{L}_u^{CE}$ &0.6361  & 0.6740 &   0.8023  \\
		\bottomrule
	\end{tabular}
\end{table}

\subsection{Ablation Study and Case Study(RQ3\&RQ4)}
In this subsection, we perform the ablation experiments and case analysis to study the impact of different parts of OptDist.  Firstly, we compare the OptDist with different derivations in terms of Norm-GINI of the overall samples: (1)\textbf{ (w/o) Gumbel-softmax}: Remove the Gumbel-softmax operation and use the plain softmax to generate the mask vector.  
(2)\textbf{ (w/o) $\mathcal{L}^{KL}$}: Remove the term of KL divergence loss from the alignment mechanism. 
 (3)\textbf{ (w/o)  $\mathcal{L}_u^{CE}$}: Remove the term of cross-entropy from the alignment mechanism. 
 (4)\textbf{ (w/o) $\mathcal{L}^{KL}$ + $\mathcal{L}_u^{CE}$}: Omit the alignment mechanism of DSM.

We then summarize the results of ablation experiments in Table ~\ref{ablation_results}. Firstly, it indicates the Gumbel-softmax operation can help the OptDist improve prediction performance. Note that Gumbel-softmax is used to achieve an approximate discrete sampling and makes each SDN focus on learning from a subset of users with similar distributions. Secondly, after removing the alignment mechanism, the performance of OptDist degraded, which indicates that the alignment mechanism can effectively alleviate the training difficulty caused by the large search space. We also conduct an ablation study on both terms to investigate further the effect of KL loss and CE loss in the alignment mechanism. As it indicates, both KL divergence loss and cross-entropy loss in the alignment mechanism boost the performance, verifying our design on soft labels and hard labels. Specifically, the cross-entropy loss can make DSM training based on a guide of the best sub-distribution, and KL divergence loss ensures the DSM also takes other sub-distribution into account.

\begin{figure}[t]
  \centering
   \subfloat[Statistic of actual CLTV ]{
    \includegraphics[width=0.48\columnwidth]{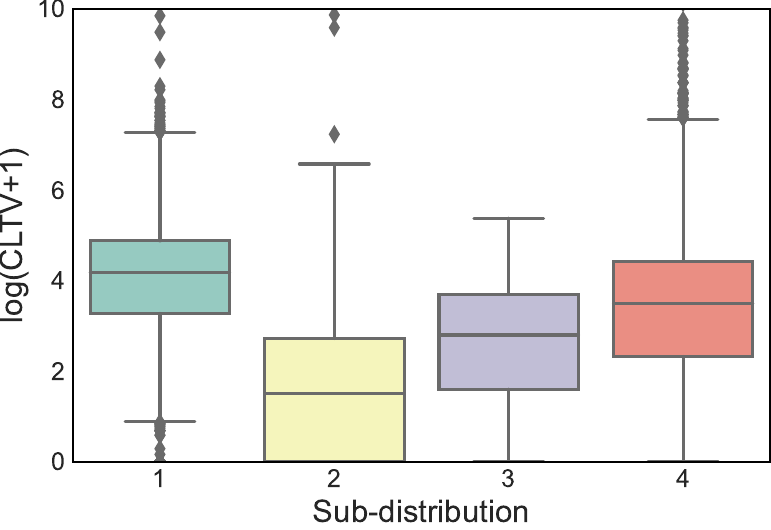}

  }
  \subfloat[Statistic of Optdist prediction]{
    \includegraphics[width=0.48\columnwidth]{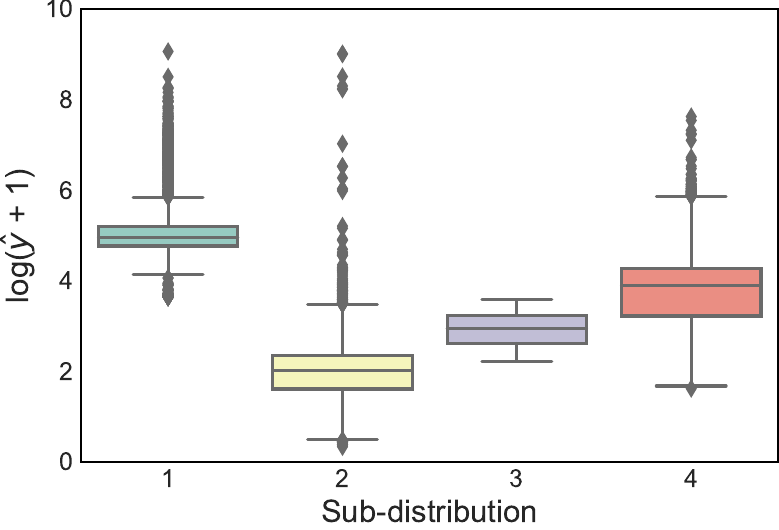}

  }
  \caption{ Difference among sub-distributions in terms of actual CLTV and predictions ($\hat{y}$) on Kaggle dataset. }
  \Description{abt_res}
  \label{vis_distribution}
\end{figure}

To intuitively illustrate the effectiveness of decomposing the distribution into multiple sub-distribution modeling in DLM, we conduct a case analysis on the Kaggle dataset. Fig.~\ref{vis_distribution} visualizes the distributions of users in each sub-distribution in terms of actual CLTV and prediction by OptDist. The box plots indicate that OptDist can select the optimal distribution for each user and fit the sub-distribution, respectively. 

\subsection{Online A/B Testing(RQ1)}
We have deployed the OptDist proposed in this paper on a large-scale financial platform to predict user CLTV on the platform and apply it to audience targeting in marketing campaigns.  

To ensure the fairness of the experiment, for each marketing campaign, we randomly take 50\% of the traffic to the experimental group and the other 50\% to the control group, ensuring that the two groups of users are homogeneous. Additionally, the marketing resources allocated to each group of traffic are equal. Following that, different models predict and rank the allocated potential users, and then select an equal number of target users for marketing campaigns.

%\begin{figure}[h]
%  \centering
%  \includegraphics[width=\linewidth]{./figures/online_pip.pdf}
%  \caption{
%Workflow of applying CLTV prediction model for online marketing %campaigns. }
%  \Description{abt_res}
%  \label{workflow}
%\end{figure}

Based on the aforementioned online A/B testing setup, we conducted online experiments on multiple marketing campaigns, focusing on users who had visited the platform in the past but had not made any purchases. The evaluation metric is the ROI, which is the ratio of the revenue contributed by users to the spend marketing budget. Table~\ref{ROI} presents the online experimental results on the large-scale financial technology platform. For each marketing campaign, we separately observe the relative improvement in ROI after 7 days, 14 days, and 30 days. The online experimental results demonstrated a significant improvement in OptDist across all marketing campaigns and observation time windows. These findings indicate the effectiveness of OptDist in real-world customer acquisition scenarios by accurately estimating the CLTV.

\begin{table}
  \caption{The relative improvement of our OptDist compared to baseline in terms of the ROI on different online campaigns.}
  \label{ROI}
\begin{threeparttable}
  \begin{tabular}{ccccc}
    \toprule
    Campaign ID& ROI-7 & ROI-14 & ROI-30   \\
    \midrule
    Campaign A & +8.96\% & +17.31\% & +21.90\%  \\
    Campaign B & +9.04\% & +9.83\% & +12.51\%\\
    Campaign C & +6.47\% & +8.68\% & +11.45\% \\
    Campaign D & +14.42\% & +16.53\% & +19.06\% \\
  \bottomrule
\end{tabular}

\end{threeparttable}
\end{table}
\section{Conclusion}
Accurately predicting CLTV is essential for increasing a company's revenue. In this paper, we propose a novel framework, OptDist, for CLTV prediction. OptDist learns multiple candidate probabilistic distributions in the DLM and adopts a network with Gumbel-softmax operation to generate exploring weights of each candidate distribution in DSM.  Additionally, we propose an alignment mechanism that generates pseudo labels for DSM according to the losses of SDNs in the DLM and uses them to guide the training of DSM, thus making the optimization more effective. In this manner, OptDist decomposes the complex single distribution modeling problem into several relatively easier-to-learn sub-distribution modeling problems and selects the optimal sub-distribution for each user. We conducted comprehensive offline experiments on two public datasets and an industrial dataset, which demonstrated the superiority of OptDist. Furthermore, we have deployed our OptDist in real-world applications and conducted online experiments in multiple marketing campaigns, and the results consistently indicated the effectiveness of OptDist.
\begin{acks}
We thank the support of the National Natural Science Foundation of China (No.62302310).
\end{acks}
\clearpage
\bibliographystyle{ACM-Reference-Format}
\bibliography{reference.bib}

\end{document}